\documentstyle[preprint,aps]{revtex}
\input epsf.tex
\def\DESepsf(#1 width #2){\epsfxsize=#2 \epsfbox{#1}}
\begin{document}
\preprint{\vbox{\hbox{}}}
\draft
\title{Implications for $B \to \eta K$ and $B \to {\rm Glueball} + K$ Modes from \\ 
Observed Large  $B\to \eta' K+X$ }
\author{$^1$ Xiao-Gang He, $^2$ Wei-Shu Hou and $^3$ C.-S. Huang}
\address{$^1$School of Physics,
University of Melbourne, Melbourne\\
$^2$ Department of Physics,
National Taiwan University,
Taipei \\ 
and \\
$^3$ Institute of Theoretical Physics,
Academia Sinica, Beijing} 
\date{December 1997} 
\maketitle
\begin{abstract}
The unexpectedly large branching ratios for 
$B\rightarrow \eta' K\ (\eta' X_s)$ decays could be of gluonic origin. 
We study the implications for 
$B\rightarrow \eta K\ (\eta X_s)$ and $P K\ (PX_s)$,
where $P$ is the pseudoscalar glueball.
In the mechanism proposed by Fritzsch,
large branching ratios are predicted for these modes. 
The $B\rightarrow \eta K$ rate is 
barely within the experimental limit,
and $B\rightarrow P K$, $PX_s$ could be at
the 0.1\% and 1\% level, respectively.
Smaller but less definite results are found for the mechanism
of $g^* \to \eta' g$ via the gluon anomaly.
\end{abstract}

Large branching ratios for exclusive $B\rightarrow \eta' K$ 
and semi-inclusive $B\rightarrow \eta' X_s$ decays 
have been reported recently, giving \cite{exp}
\begin{eqnarray}
Br(B^\pm \rightarrow \eta' K^\pm) &=& 
(7.1^{+2.5}_{-2.1}\pm 0.9)
\times 10^{-5}\;,\nonumber\\
Br(B^0\rightarrow \eta' K^0)&=&
(5.3^{+2.8}_{-2.2}\pm 1.2)\times 10^{-5}\;,\nonumber\\
Br(B\rightarrow \eta' X_s) &=& (6.2\pm 1.6\pm 1.3)\times 10^{-4}\;\;\;\;
\mbox{($2.0 < p_{\eta'} < 2.7$ GeV)}\;.
\end{eqnarray}
Factorization calculations of  four quark operators in the Standard 
Model (SM) indicate that exclusive branching ratios 
could be accounted for by choice of form factors \cite{dhp,ab,ct}. 
However, the four quark operators do not seem
sufficient for the semi-inclusive decay \cite{dhp}.  
Several mechanisms have been proposed \cite{as,ht,hz,fri,kp,more}
to explain the latter, and
ways to distinguish some of the mechanisms have also been proposed\cite{dy}.
We wish to study here the implications for 
gluonic origins \cite{as,ht,fri} of such enhancements.
In particular, we concentrate on the mechanism proposed in Ref. \cite{fri},
which postulates an {\it ad hoc} effective Hamiltonian of the form
\begin{eqnarray}
H_{eff} = a\;  \alpha_s G_F\, \bar s_L b_R G_{\mu\nu} \tilde G^{\mu\nu}.
\label{hamilt}
\end{eqnarray}
In the notation of Ref. \cite{fri}, $a$ contains a factor of $V_{ts}$.
Similar effective Hamiltonians can be 
generated perturbatively at one loop level in the SM\cite{sm-gg}.
However, the resulting $Br(B\rightarrow \eta' X_s)$ would be 
smaller by more than two orders of magnitude. 
It was therefore argued in Ref. \cite{fri} that Eq. (2) 
may arise from non-perturbative effects. 
It may also arise from new physics. 
It turns out that a single parameter $a$
could account for both inclusive and exclusive $\eta'$ rates of Eq. (1).
Although the semi-inclusive $m_{X_s}$ spectrum seems to 
favor 3-body decay over 2-body,
which supports the anomaly induced $g^*\to \eta' g$ mechanism \cite{as,ht},
some smearing effect cannot yet be ruled out \cite{fri}.
Fitting Eq. (2) to Eq. (1), we find that $\eta$
and pseudoscalar glueball $P$ modes could then be quite large:
$Br(B\rightarrow \eta K)$ is barely within the experimental limit,
while $B\rightarrow P K,\ P X_s$ could be at the
0.1\% and 1\% level, respectively.

To calculate $Br(B\rightarrow \eta' X_s)$ from Eq. (2), one has to project out the
$G_{\mu\nu} \tilde G^{\mu\nu}$ content in $\eta'$, i.e. extract the 
matrix element $\langle 0\vert G_{\mu\nu} \tilde G^{\mu\nu} \vert  \eta'\rangle$. 
Using the following definitions
\begin{eqnarray}
&&\eta= \eta_8\,  \cos \theta -\eta_0\,  \sin  \theta\;, \ \ \
\eta'= \eta_8\,  \sin  \theta + \eta_0\,  \cos \theta\;,\nonumber\\
&&\langle 0\vert j^8_\mu\vert  \eta_8\rangle  
=\langle 0\vert {\bar u\gamma_\mu \gamma_5 u + \bar d \gamma_\mu \gamma_5 d
-2\bar s \gamma_\mu \gamma_5 s\over \sqrt{6}}\vert \eta_8\rangle 
= if_8\,  p_\mu\;,\nonumber\\
&&\langle 0\vert j^0_\mu\vert  \eta_0\rangle  
=\langle 0\vert {\bar u\gamma_\mu \gamma_5 u + \bar d \gamma_\mu \gamma_5 d
+\bar s \gamma_\mu \gamma_5 s\over \sqrt{3}}\vert \eta_0\rangle 
= if_0\,  p_\mu\;,
\end{eqnarray}
we obtain \cite{coupling}
\begin{eqnarray}
\langle 0\vert \partial^\mu j^8_\mu\vert \eta\rangle  
&=& f_8\, \cos \theta\,  m_\eta^2 = {1\over \sqrt{6}}
\langle 0\vert i2m_u\bar u\gamma_5 u +i2m_d \bar d\gamma_5 d
                       - i4m_s \bar s \gamma_5s \vert \eta\rangle \;,\nonumber\\ 
\langle 0\vert \partial^\mu j^8_\mu\vert \eta'\rangle  
&=& f_8\, \sin \theta\,  m_{\eta'}^2 ={1\over \sqrt{6}}
 \langle 0\vert i2m_u\bar u\gamma_5 u +i2m_d \bar d\gamma_5 d
                        - i4m_s \bar s \gamma_5s \vert \eta'\rangle \;,\nonumber\\ 
\langle 0\vert \partial^\mu j^0_\mu\vert \eta\rangle  
&=&- f_0\,  \sin \theta\,  m_{\eta}^2 
={1\over \sqrt{3}}
 \langle 0\vert i2m_u\bar u\gamma_5 u +i2m_d \bar d\gamma_5 d
                        +i2m_s \bar s \gamma_5s 
+{3\alpha_s\over 4 \pi} G_{\mu\nu}\tilde G^{\mu\nu}\vert \eta\rangle \;,\nonumber\\ 
\langle 0\vert \partial^\mu j^0_\mu\vert \eta'\rangle  
&=& f_0\,  \cos \theta\,  m_{\eta'}^2 
={1\over \sqrt{3}}
 \langle 0\vert i2m_u\bar u\gamma_5 u +i2m_d \bar d\gamma_5 d
                        +i2m_s \bar s \gamma_5s 
+{3\alpha_s\over 4 \pi} G_{\mu\nu}\tilde G^{\mu\nu}\vert \eta'\rangle \;.
\end{eqnarray}
Neglecting the small up and down quark masses, we arrive at 
\begin{eqnarray}
\langle 0\vert {3\alpha_s\over 4\pi} G_{\mu\nu}\tilde G^{\mu\nu}\vert \eta\rangle 
&=& \sqrt{{3\over 2}}(\cos \theta f_8 -\sqrt{2}
 \sin \theta f_0)\, m_\eta^2\;,\nonumber\\
\langle 0\vert {3\alpha_s\over 4\pi} G_{\mu\nu}\tilde G^{\mu\nu}\vert \eta'\rangle 
&=& \sqrt{{3\over 2}}(\sin \theta f_8 +\sqrt{2} \cos \theta f_0)\, m_{\eta'}^2\;.
\label{glue}
\end{eqnarray}

We emphasize that Eqs. (3)--(5) do not make explicit assumptions of
the quark and gluon content of $\eta$ and $\eta'$ mesons.
The parameters $\theta$, $f_8$ and $f_0$, however, are still not quite certain.
In the following we shall use 
(A) $\theta = - 17.0^\circ$, $f_8/f_\pi \simeq f_0/f_\pi = 1.06$ \cite{coupling},  and
(B) $\theta = - 22.0^\circ$, $f_8/f_\pi = 1.38$ and $f_0/f_\pi = 1.06$ \cite{vh,pdg},
to illustrate the sensitivity to $\eta$-$\eta'$ parameter values.
To account for the observed central value of 
$Br(B\rightarrow \eta' X_s) = 6.2\times 10^{-4}$, 
one needs $a_{\rm in}\approx 0.012\ (0.015)$ GeV$^{-1}$,
where the subscript stands for inclusive.
The numbers are given for Set A (Set B) respectively, 
a notation which we shall employ from now on.

Assuming that Eq. (\ref{hamilt})
also dominates the large exclusive $Br(B\rightarrow \eta' K)$, 
one finds
\begin{eqnarray}
A(B^-\rightarrow \eta' K^-)
= a\, G_F {4\pi \over 3}\sqrt{{3\over 2}}(\sin \theta f_8 + \sqrt{2} \cos \theta f_0)
m^2_{\eta'}\, {m_B^2-m_K^2\over 2(m_b+m_s)} \,
F^K_0(m_{\eta'}^2)\;,
\end{eqnarray}
where the form factor $F^K_0(m_{\eta'}^2)$ is estimated to be 0.33 \cite{stech}. 
With the $a_{\rm in}$ values obtained earlier, 
we find $Br(B\rightarrow \eta' K) \simeq 6.6 \times 10^{-5}$, 
which is consistent with experimental data.
In turn, we could extract
$a_{ex} = 0.012\ (0.016)$ GeV$^{-1}$ from $Br(B^\pm \rightarrow \eta' K^\pm)$, and
$a_{ex} = 0.011\  (0.014)$ GeV$^{-1}$ from $Br(B_0\rightarrow \eta' K^0)$,
which are consistent with $a_{\rm in}$.

We can easily obtain $Br(B\rightarrow \eta K\, (\eta X_s))$ by using Eq. (\ref{glue}). 
If the mechanism of Ref. \cite{fri} is the sole source for the large 
$B\rightarrow \eta' K\, (\eta' X_s)$ branching ratios, then 
$Br(B\rightarrow \eta K\, (\eta X_s))$ is simply related to 
$Br(B\rightarrow \eta' K\, (\eta' X_s))$ by
\begin{eqnarray}
Br(B\rightarrow \eta K\, (\eta X_s)) =R^2\,  \Phi \,  Br(B\rightarrow \eta' K\, (\eta' X_s))\;,
\end{eqnarray}
where
\begin{eqnarray}
R = {A(B\rightarrow \eta K (\eta X_s))\over A(B\rightarrow \eta' K(\eta' X_s))}
={\cos \theta f_8 - \sqrt{2} \sin \theta f_0\over sin\theta f_8 + \sqrt{2} 
\cos \theta f_0}\,  {m_\eta^2\over m_{\eta'}^2}\;,
\end{eqnarray}
and  $\Phi $ is a phase space correction factor. 
For inclusive decays,
\begin{eqnarray}
\Phi  = {\sqrt{(1-(m_\eta+m_s)^2/m_b^2)(1-(m_\eta-m_s)^2/m_b^2)}
\over \sqrt{(1-(m_{\eta'}+m_s)^2/m_b^2)(1-(m_{\eta'}-m_s)^2/m_b^2)}}\;,
\end{eqnarray}
while for exclusive decays

\begin{eqnarray}
\Phi  = {\sqrt{(1-(m_\eta+m_K)^2/m_B^2)(1-(m_\eta-m_K)^2/m_B^2)}
\over \sqrt{(1-(m_{\eta'}+m_K)^2/m_B^2)(1-(m_{\eta'}-m_K)^2/m_B^2)}}\;.
\end{eqnarray}
Note that Eqs. (5) and (8) give $R \simeq$  0.42 (0.69). 
This means that the slight numerical change from Set A to Set B
results in a factor of 2.7 difference in predictions for $\eta$ modes. 
The branching ratio for inclusive $B\rightarrow \eta X_s$ is therefore
\begin{eqnarray}
Br(B\rightarrow \eta X_s) = 1.1\ (3.0) \times 10^{-4}\  (p_\eta > 2.1 \ 
\mbox{GeV})\;,
\end{eqnarray}
which is still consistent with the experimental upper limit of $ 4.4\times 10^{-4}$
at the 90\% C.L. for $2.1\ \mbox{GeV} < p_\eta < 2.7\ \mbox{GeV}$.
For exclusive decays,
using $a_{\rm in}$ from above gives 
\begin{eqnarray}
Br(B^-\rightarrow \eta K^-) = 1.2\ (3.2) \times 10^{-5}\;,
\end{eqnarray}
which is higher than the experimental limit of $0.8\times 10^{-5}$ (90\% C.L.).
If Set B is the case, then the Fritzsch scenario of Eq. (2) would be in 
grave difficulty.
Scaling from exclusive $B\rightarrow \eta' K$ decay gives similar results.
It is somewhat surprising that the one parameter model of Fritzsch, Eq. (2),
could account for both $B\to \eta'$ and $B\to \eta$ data.
Our results underpin the importance of 
improved experimental measurements in probing this mechanism.

It is interesting to speculate that,
since this mechanism invokes the large gluon content of $\eta'$,
there should be a large branching ratio for $B\rightarrow P K$,
where $P$ is the pseudoscalar glueball.
To obtain the branching ratio for $B\rightarrow P K$,
we need to know the matrix element
$\langle 0\vert \alpha_s G_{\mu\nu}\tilde G^{\mu\nu}\vert P\rangle $.
This can be obtained by using QCD sum rules, as in Refs. \cite{glueball,mass}.
Early calculations give $(3/4\pi)\langle 0|\alpha_s 
G_{\mu\nu}^a\tilde G^a_{\mu\nu}|P\rangle
\approx 0.87 f_\pi\, m_P^2$ for glueball mass around 1.4 GeV (i.e. taking 
$\eta(1440)$ as glueball candidate). Recent lattice
calculations suggest, however, that glueball masses are heavier,
i.e.  $m_P = 2.3\pm 0.2$ GeV\cite{lattice},
and the sum rule result could be quite different.
We will follow the analysis in Ref. \cite{mass}, but include the $\eta$ contribution
to the low energy spectral density which was previously neglected.

The basic idea of QCD sum rule analysis in the present case is to 
match the dispersion relation for the vacuum topological susceptibility $T(s)$
with the 
result found by using the operator product expansion. 
The definitions of $T$ and the spectral density $\rho$ are given by
\begin{eqnarray}
T(s) &=& i\int d^4x e^{iqx}
\langle 0|T(j_{ps}(x) j_{ps}(0))|0\rangle\;,\;\;
j_{ps}= {3\alpha_s\over 4\pi} G^a_{\mu\nu} \tilde G_{\mu\nu}^a\;,\nonumber\\
\rho(s) &=& \rho^{pole}(s) + \rho^{cont}(s)\, \theta(s-s_1)\;,\nonumber\\
\rho^{pole}(s)& =& \sum_i f_i^2\, m_i^4\, \delta (s-m_i^2)\;,\;\;
\rho^{cont}(s)\simeq \left ( {3\alpha_s\over 4\pi}\right )^2 
                                     {2\over \pi}\left (1 + 5{\alpha_s\over \pi}\right ) s^2
                                     \equiv b s^2 ,\nonumber\\
f_i\, m_i^2&=& \langle 0|{3\alpha_s\over 4\pi} G_a^{\mu\nu} \tilde G_a^{\mu\nu} | i\rangle\;,
\end{eqnarray}
where $i= P, \eta',\eta$, and $s_1$ is the continuum threshold.
To improve series convergence, 
and to remove subtraction dependence,
one makes the following Borel transformations on $T(s)$:
$\int_0^\infty {\rm Im\,} T(s)\, e^{-s/M^2} ds/s$ and
$\int^\infty_0 {\rm Im\,} T(s)\, e^{-s/M^2}ds$, with $M^2$ a free parameter. 
One gets
\begin{eqnarray}
\int^\infty_0 {e^{-s/M^2}\over s}\rho^{pole}(s) ds &=& \int^{s_1}_0 {e^{-s/M^2}\over s}
\rho^{cont}(s) ds + \pi \left ( {3\alpha_s \over 4\pi}\right )^2 ( D_4 +{D_6\over M^2} +...)
\;,\nonumber\\
\int^\infty_0 e^{-s/M^2}\rho^{pole}(s) ds &=& \int^{s_1}_0 e^{-s/M^2}
\rho^{cont}(s) ds + \pi \left ( {3\alpha_s \over 4\pi}\right )^2 (D_6
+  O({1\over M^2}) +...)\;,
\end{eqnarray}
where
$D_4 = 4\, \langle 0|G^a_{\mu\nu}G_{\mu\nu}^a|0 \rangle$ and 
$D_6 = 8g_s f^{abc}\langle 0| G^a_{\mu\alpha}G^b_{\alpha\beta}G^c_{\beta\mu}|0\rangle$.
Numerically we use 
$\tilde D_{4,6} \equiv \pi(3\alpha_s/4\pi)^2 D_{4,6}
 = (1.44\pm 0.61) \times 10^{-2}$ GeV$^4$, 
$ (0.25\pm 0.11)\times 10^{-2}$ GeV$^6$ \cite{mass} respectively.
%
To further simplify, we follow
Ref. \cite{mass} and take $M^2$ to infinity, leading to
\begin{eqnarray}
f_p^2 m_P^2 + f_\eta^2m_\eta^2 + f_{\eta'}^2m_{\eta'}^2
 &=& {bs_1^2\over 2} + \tilde D_4,\nonumber\\
f_p^2 m_P^4 + f_\eta^2m_\eta^4 + f_{\eta'}^2m_{\eta'}^4
 &=& {bs_1^3\over 3 } + \tilde D_6.
\end{eqnarray}
Since $f_{\eta,\eta'}$ and $m_{\eta,\eta'}$ are known, 
$f_P$ and $m_P$ can be determined from knowing what to take
for the continuum threshold $s_1$. 

Some information on $s_1$ can be obtained from comparing with lattice studies,
which give $m_P = 2.3 \pm 0.2 $ GeV 
in the quenched approximation.
The quenched approximation is equivalent to QCD without quarks
(but lattice studies incorporate running coupling with $N_f = 3$). 
Applying QCD sum rules, we obtain similar results as in Eq. (15), 
except that there are no $\eta$ and $\eta'$ mesons. 
Also, in the absence of quarks, 
there are no light meson states for glueball to decay into, 
hence the continuum threshold $s_0$ in this case would be larger than $s_1$.
Using the lattice glueball mass of 2.3 GeV, 
$s_0$ is determined to be $7.4$ GeV$^2$\cite{mass}.
For the lower bound on $s_1$, we require it to be larger than the
glueball mass, since otherwise the sum rule approach breaks down.
The glueball mass in QCD need not be the same as 
the lattice result in quenched approximation.
To get an idea on how $m_P$ and $f_P$ can be determined, 
we take $s_1$ to range from 7.4 GeV$^2$ to 6 GeV$^2$.
Using the two sets of $\theta$ and $f_{0,8}$ discussed earlier 
and $\alpha_s = 0.33$ (at $m_P$ scale),
we get 
\begin{eqnarray}
m_P &=& 2.42\ \mbox{GeV},\ f_P = 1.15\ (1.16)f_\pi, \ \ \ \ 
                            \mbox{for}\ s_1 = 7.4\ \mbox{GeV}^2 \nonumber \\
m_P &=& 2.29\ \mbox{GeV}, \ f_P = 0.92\ (0.94) f_\pi, \ \ \ \ 
                            \mbox{for}\ s_1 = 6.0\ \mbox{GeV}^2.
\label{num}
\end{eqnarray}
The difference between Set A and B is small in the given range for $s_1$.
In Ref.\cite{mass} the lower limit for $s_1$ was
 taken to be as low as 3 GeV$^2$.
We consider this to be too low for the following reason. 
The effect of including $\eta$ in the sum rule is
not significant for $s_1$ between 7.4 to 6.0 GeV$^2$,
but cannot be ignored for $s_1 \simeq$ 3 GeV$^2$. 
The inclusion of $\eta$ increases the value for $s_1$ 
such that $s_1$ is larger than 3.24 GeV$^2$ (3.15 GeV$^2$) for
Set A (Set B), and below these values 
there is no physical solution to the sum rule equations.
When $s_1$ is close to the above lower bounds,
the solution is in any case very sensitive to small changes.

We are now ready to calculate the branching ratio for $B\rightarrow P K\, (P X_s)$.
Using the glueball masses and decay constants in Eq. (\ref{num}), we obtain
\begin{equation}
{Br(B\rightarrow P K\, (PX_s))\over Br(B\rightarrow \eta' K\, (\eta' X_s))} \simeq
\left \{ \begin{array}{ll}
1.9 \ (3.2 )  \;\ \ \ \mbox{for}\ m_P = 1.44\ \mbox{GeV}\\
\, 11\,\ (20) \,\ \ \ \ \mbox{for}\ m_P = 2.29\ \mbox{GeV}\\
\, 21\,\ (36) \,\ \ \ \ \mbox{for}\  m_P = 2.42\ \mbox{GeV}
\end{array}
\right .\end{equation}
As argued, the low glueball mass value of $1.44$ GeV is not plausible.
The numerical value for this case is taken from Ref. \cite{glueball} 
for sake of illustration.
The branching ratio is very sensitive to $m_P$ because 
$\langle 0\vert (3\alpha_s/4\pi) G_{\mu\nu}\tilde G^{\mu\nu}\vert P\rangle
\propto m_P^2$.
Hence, $Br(B\rightarrow  P X_s)$ is at least twice as
large as $Br(B\rightarrow \eta' K(\eta' X_s))$, 
and is likely as large as 1\%, which is truly huge.
Since the perturbative estimate for the inclusive $b\to sg^*$ penguin
is itself $\simeq 1\%$ \cite{HSS}, this indicates once again that the Fritzsch
mechanism is probably not the sole source for the large branching ratios
for $B\rightarrow \eta' K\, (\eta' X_s)$.
However,  it does suggest that one should 
search for the pseudoscalar glueball in $B$ decays,
since the scenario is not yet completely ruled out.
For exclusive decay, one finds $Br(B^\pm \rightarrow  P K^\pm) \sim$ 0.1\% 
or even higher for $m_P \simeq 2.3$ GeV.
Taking $Br(\eta_c\to K\bar K\pi ) \simeq 6.6\% $ as a guide, 
one should search for a fast $K^\pm$ ($p_K \simeq 2.1$ GeV) recoiling against 
a $K\bar K \pi$ system with mass $\sim$ 2.2--2.3 GeV.
Background should be small since both $B\to D_sK +X$ and 
usual $D\to K\bar K + X$ decays are suppressed.
Such a large rate for $B\to PK$ should be readily observable
if $\Gamma_P$ is not too broad,
which is likely the case since the $2^{++}$ glueball
candidate, the $\xi(2220)$, has a width of only 20 MeV \cite{xi}.

Before we conclude, let us comment on the predictions for 
$B\rightarrow \eta X_s$ and $P X_s$ for the mechanism proposed in Refs.\cite{as,ht}.
It was proposed in  Ref. \cite{as} that the 
coupling of $\eta'$ to two gluons via the gluon anomaly 
may be able to explain the semi-inclusive decay by the process
$b\rightarrow s g^* \rightarrow s g \eta'$ within the SM. 
However, it was subsequently pointed out that,
because the effective $\eta'$-$g$-$g$ coupling arises from the 
Wess-Zumino term \cite{coupling,gge},
\begin{eqnarray}
L_{eff} = \alpha_s{\sqrt{N_F}\over 4\pi f_0}\, \eta_0 
G_{\mu\nu} \tilde G^{\mu\nu}\;,
\label{amp}
\end{eqnarray}
it is proportional to $\alpha_s$. If this runs with
the $q^2$ of the virtual gluon $g^*$,  the branching 
ratio is reduced by a factor of 3\cite{ht}. 
Enhanced $bsg$ coupling due to new physics is then needed \cite{ht},
with the tantalizing prospect of potentially large CP violation effects. 
A critical question is the form factor behavior of Eq. (\ref{amp}).
Analogy with the photon anomaly suggests $1/k^2$ suppression \cite{kp},
where $k$ is the difference between the two gluon momenta.
However, the analogy fails precisely \cite{ht} because of
binding effects in the two gluon channel,
with the normalization of  Eq. (\ref{amp}) fixed by the $\eta$-$\eta'$
system at low energy.
The study of the off-shell behavior of the gluon anomaly has yet to begin.

Turning back to phenomenology, 
since one of the gluons appear in the final state, 
one can no longer use Eq. (5).
Using naive $\eta$--$\eta'$ mixing as an estimate of
the ratio 
%
 $A(b\rightarrow s g \eta)/A(b\rightarrow s g \eta') \sim \tan \theta$,
the inclusive $Br(B\rightarrow \eta X_s)$ could well be below the
experimental upper limit.
For the exclusive process,
as the actual transition is $b\bar q \to \eta' + sg\bar q$ 
and there are no IR singularities, 
it was noted \cite{ht} that  if
the $sg\bar q$ system evolves into the $K$ meson when the
gluon is soft, it could account for the exclusive $B \to \eta' K$ rate.
However, since it is difficult to make quantitative predictions,
the large exclusive $B\rightarrow \eta' K$ branching ratio 
could well be coming from other mechanisms, 
such as from four quark operators in the SM with appropriate
form factors.

To evaluate the branching ratio for $B\rightarrow P X_s$, 
we need to estimate the $Pgg$ coupling. 
For sake of illustration we use an Ansatz given in Ref. \cite{gounaris},
\begin{eqnarray}
L_{eff} = - {2c\Lambda\, \over k^2} P\, G_{\mu\nu} \tilde G^{\mu\nu}\;,
\label{glue1}
\end{eqnarray}
where $k$ is the difference of the two gluon momenta. 
For $g^* \to Pg$ where $P$ and $g$ are both on-shell,
one has $k^2 = 2q^2 -m_P^2$ where $q$ is the virtual gluon momentum.
But in the small  $k^2$ limit, Eq. (\ref{glue1})
could generate finite glueball mass \cite{gounaris} at one loop level, giving
\begin{eqnarray}
m_P^2 = {16 c^2\Lambda^2\over \pi^2} (4 \ln 2-1)\;.
\end{eqnarray}
Of course there should be other contributions to $m_P^2$,
but saturating $m_P^2$ with the above relation gives the upper bound
$(c_1\Lambda )_{max} = 0.59\, m_P$,
and Eq. (\ref{glue1}) becomes
\begin{eqnarray}
L_{eff} \simeq  - {1.2\, m_P^2 \over 2q^2 - m_P^2} \,
                          {P\over m_P}\, G_{\mu\nu} \tilde G^{\mu\nu}\;.
\end{eqnarray}
Comparing with Eq. (\ref{amp}),
the most salient feature is the absence of an $\alpha_s$ factor.
It also highlights our ignorance of the form factor behavior of Eq. (\ref{amp}).
We emphasize that $m_{\eta'}$ does not arise from $\eta'$-$P$ mixing,
nor from some Ansatz analogous to Eq. (19),
but from topological effects through the gluon anomaly \cite{gge}.
%
Since the effective $P$-$g$-$g$ coupling of Eq. (21) is only slightly larger than
the $\eta'$-$g$-$g$ coupling of Eq. (18), we find 
$Br(B\to P X_s) \ \raisebox{-.5ex}{\rlap{$\sim$}} \raisebox{.4ex}{$>$}\ 
Br(B\to \eta' X_s)$ is likely.
Though sizable, the branching ratios for $B\rightarrow P  X_s$ and $PK$ 
in the scenario of Refs. \cite{as,ht} are not as large as in the case of Ref. \cite{fri}.
However, 
more quantitative predictions in the present case are difficult,
especially for the exclusive case.

To conclude, we have shown that the mechanism proposed by Fritzsch for 
explaining large $Br(B\rightarrow \eta' K\ (\eta' X_s))$
gives large branching ratios for $B$ decays with an $\eta$ or a glueball
in the final state. The predicted branching ratio for $B\rightarrow \eta K$
is barely within the experimental upper limit,
while $Br(B\rightarrow P K)$ could be as large as 0.1\%,
with $Br(B\to P X_s)$ ten times larger. 
If this is the case, perhaps the pseudoscalar glueball $P$ 
could be first discovered in $B$ decays.
Improved experiments will provide decisive information about this mechanism.

\acknowledgements

This work was supported in part by the ARC, 
the NSC, 
and the NNSF.
XGH acknowledges the support of K.C. Wong Education Foundation, Hong Kong.


%
%
%
%

\end{document}